# Correlating Cell Behavior with Tissue Topology in Embryonic Epithelia

Sebastian A. Sandersius[1], Manli Chuai[2], Cornelis J. Weijer[2], Timothy J. Newman[1,2]*

1 Department of Physics, Center for Biological Physics, Arizona State University, Tempe, Arizona, United States of America, 2 Division of Cell and Developmental Biology, Wellcome Trust Biocentre, College of Life Sciences, University of Dundee, Dundee, United Kingdom

## Abstract

Measurements on embryonic epithelial tissues in a diverse range of organisms have shown that the statistics of cell neighbor numbers are universal in tissues where cell proliferation is the primary cell activity. Highly simplified non-spatial models of proliferation are claimed to accurately reproduce these statistics. Using a systematic critical analysis, we show that non-spatial models are not capable of robustly describing the universal statistics observed in proliferating epithelia, indicating strong spatial correlations between cells. Furthermore we show that spatial simulations using the Subcellular Element Model are able to robustly reproduce the universal histogram. In addition these simulations are able to unify ostensibly divergent experimental data in the literature. We also analyze cell neighbor statistics in early stages of chick embryo development in which cell behaviors other than proliferation are important. We find from experimental observation that cell neighbor statistics in the primitive streak region, where cell motility and ingression are also important, show a much broader distribution. A non-spatial Markov process model provides excellent agreement with this broader histogram indicating that cells in the primitive streak may have significantly weaker spatial correlations. These findings show that cell neighbor statistics provide a potentially useful signature of collective cell behavior.

**Funding:** This work was supported by a grant from the Human Frontiers Science Program (RGP-0038, www.hfsp.org) and the National Institutes of Health Physical Sciences Oncology Centers (5U54CA143862, www.nih.gov). The funders had no role in study design, data collection and analysis, decision to publish, or preparation of the manuscript.

**Competing Interests:** The authors have declared that no competing interests exist.

* E-mail: t.newman@dundee.ac.uk

## Introduction

Development of higher organisms is dependent on extensive division and movement of cells arranged in well-organized, densely packed epithelial sheets, in many instances only one cell layer thick. During early development cell divisions may be synchronous, but later in development divisions tend to become asynchronous and at any given instant in time only a small fraction of cells in the sheet will be dividing. Likewise, cell dynamics (e.g. movement within and out of the plane of the epithelial sheet) may have varying degrees of coherence or cooperativity. Understanding the control of growth and spatial organization of these embryonic epithelial sheets is a central goal of the study of development. Major questions concern how tissue structure and dynamics are driven by individual cell behaviors; for example, whether the axes of cell divisions are organized on a tissue-wide scale, possibly resulting in directional tissue elongation, and whether there is any logic as to which cells become neighbors after division [1]. Cells in these embryonic epithelial sheets often have approximately polygonal cross-sections in the plane of the sheet. It would be expected that in cases where there is little proliferation or autonomous cell dynamics the packing of the cells would be close to optimal hexagonal packing. Conversely, it is well established that in proliferating epithelia, as commonly found in developing embryos, the polygonal nature of the cells is significantly more diverse. The statistics describing this diversity of cell neighbor numbers (CNN) have recently been observed to be strikingly universal across diverse taxa [2].

A useful way to characterize this "tissue topology" is to construct a histogram of CNN. Note, assuming that cells have well-defined sides, the number of sides of a given cell is equal to its CNN (i.e. number of nearest neighbors). Studies measuring such histograms date back to the 1920's with analysis of proliferating epidermis in cucumber [3,4]. More recently CNN histograms have been measured for proliferating epithelial tissues in *Drosophila*, *Hydra*, *Xenopus*, *Anagallis* and *Arabidopsis* organisms [2,5–8]. In particular, Gibson et al. (2006) [2] (hereafter referred to as GPNP) measured the CNN histograms in three diverse model organisms: *Drosophila* (fruit fly), *Xenopus* (frog), and *Hydra* (marine invertebrate), and observed that these histograms fall approximately onto a "universal" curve. The CNN histograms for the plants mentioned above also fall approximately onto the universal curve [7]. GPNP found that cells with six nearest neighbors were the most common, but significant numbers of cells with five or seven nearest neighbors were also counted. The authors were able to reproduce the histogram with surprisingly good precision using a non-spatial Markov chain model. There appeared to be one discrepancy between calculations and observations, i.e. a small but significant number of 4-sided cells was observed experimentally, but was absent in the computational histograms derived using the Markov chain model. Such non-spatial models ignore the spatial correlations between CNN of nearby cells, which at first glance would appear to be a dramatic over-simplification.

In this paper we study proliferating epithelia in the chick embryo. This allows the study of the effect on geometric order of additional cell behaviors; namely, movement within the sheet and out of the sheet (ingression). The early chick embryo has the form of an epithelial-like sheet (the epiblast) in which massive collective cell movement occurs during the process of primitive streak formation. The primitive streak demarcates a region to which cells in the epiblast migrate and then ingress into the space below, to eventually form mesoderm and endoderm tissues [9]. Cells in the region of the streak are believed to undergo a process not dissimilar to EMT (epithelial to mesenchymal transition). The epithelial-like tissue in the epiblast is single-layered, except for the primitive streak which has a multi-layered structure. We collected data of CNN from the chick embryo in which three distinct tissue phenotypes can be studied within a relatively short time interval marked by formation of the primitive streak. Proliferation is common to all three phenotypes occurring in the chick epiblast. In the pre-streak (Pre-S) tissue, there is little or no cell migration or ingression. A few hours later in development the streak begins to form. Lateral to the streak (LS), cell movement is locally coherent, meaning cells are collectively migrating, but retain the same neighbors for significant time periods, and cell morphologies do not appear significantly distorted. Within the streak (WS), cell morphologies are significantly distorted and movement appears to be locally incoherent due in part to cells unilaterally ingressing beneath the epiblast. This distinction between locally incoherent and coherent dynamics is important for what follows. The histogram obtained for Pre-S is narrow, and agrees well with the universal histogram measured by GPNP. The histogram for LS is also narrow and within error bars is consistent with the universal histogram. Conversely, the histogram for WS is far broader, with a long tail indicating cells with as many as 11 or 12 neighbors.

We use an array of modeling techniques to interpret our data. We start by revisiting the non-spatial Markov chain model of GPNP. Our attempts to improve the biological realism of this model (allowing transient 3-sided cells, investigating the implementation of a random division axis, and reformulating the model as a Markov process to represent asynchronous cell division) all lead to *significant deviations* from the universal histogram. This indicates that a non-spatial model is not able to robustly describe the histogram of CNN, and that spatial correlations must be accounted for in describing tissue with locally coherent dynamics. This is consistent with the intuition gained from collective behavior in physical systems (e.g. magnetic systems near to the critical point [10]), where spatial correlations are central to the quantitative understanding of the system statistics. Thus, we implement a fully spatial model by simulating the growth of a proliferating epithelium using the recently developed Subcellular Element Model (ScEM) [11]. This model incorporates both asynchronous cell division and explicit spatial arrangement of cells. Our spatial simulations produce histograms which are in good agreement with those measured experimentally. We test the robustness of these histograms by varying the criterion for what constitutes a cell side (i.e. how finely one resolves apparent four-way junctions). We find that by varying a cut-off parameter, our simulations can produce histograms which interpolate between those measured by GPNP, and significantly different histograms which were measured by Farhadifar et al. (2007) [5] for the *Drosophila* imaginal wing disk. A number of spatially explicit models have recently been applied to CNN histograms [5–7,12–14]. We defer a discussion of their relative merits, and their relation to the ScEM simulations, until later in the paper.

To produce a spatial model of WS, in which one has cell proliferation, movement, and ingression, requires significant extensions of the ScEM, and is beyond the scope of this paper. However, given that the cell dynamics in this region is less spatially coherent than in Pre-S and LS (i.e. ingression will tend to break up spatial correlations) one might argue that a non-spatial model would have some utility in this case. Remarkably, we find that our non-spatial Markov process model, which ignores spatial correlations but accounts for asynchronous cell division, provides very satisfactory agreement with the histogram for WS.

## Results

### Experimental histograms

We have measured the distribution of neighbors in the epiblast of the early chick embryo at stage EGXII of development [15], which is pre-streak (Pre-S), as well as at stage HH2 where there is already a significant amount of cell movement occurring associated with the formation of the primitive streak [16]. File S1 contains more details of the data analysis. At stage HH2 we measured distributions for cell populations both lateral to the streak (LS) and within the region of the streak (WS). The epiblast of the EGXII stage chick embryo, which will give rise to all the cells of the embryo proper, forms a highly polarized epithelial sheet where cells form tight and adherens junctions at their apical site and contact a basal lamina on their basal side. Cell divisions in the epiblast of the chick embryo, at this stage of development, occur essentially randomly in space and time ($\sim 4\%$ of the cells appear to be in mitosis at any given time), and there is still little cell movement taking place. Cells in the epiblast round up during cell division every 8–10 hours, form a mitotic plane in an apparently random direction, and then resolve into two cells that regain their columnar morphology in the epiblast. To outline the cells for the Pre-S measurements we have stained the apical cell boundaries with fluorescently labelled rhodamine phalloidin that specifically binds filamentous actin which is highly enriched in the cell cortex and thus outlines the cell shape. Confocal images were generated and maximum projection Z stack analyzed (Figure 1a). At stage HH2 of development, the primitive streak has begun to bisect the epiblast. Cells lateral to the streak are undergoing coherent cell movement within the plane of the epithelial sheet. Cells within the streak itself ingress to a layer beneath the epiblast by undergoing an EMT. This allows them to become progressively detached from their neighbors, as they shift their mass away from the apical surface of the epithelial sheet. For LS and WS measurements, we used antibodies against the tight junction component ZO1 in the epiblast of fixed embryos at the HH2 stage (Figures 1b and 1c). Ingressing cells in WS can be identified by their unusually small apical areas in Figure 1c. CNNs were counted manually to generate histograms (Figure 2a). It is seen that the distributions for Pre-S and LS are quite similar, whereas the distribution for WS is considerably broader, with a long tail extending to CNNs greater than 10. In Figure 2b we plot the Pre-S histogram against the experimental histograms obtained by a number of groups for proliferating epithelia in diverse embryonic systems, and note that the chick Pre-S histogram falls neatly onto the universal histogram.

### Modeling the universal histogram

**Markov chain models.** GPNP describe the universal histogram using a non-spatial model of cell division which assumes synchronous cell division, allowing a Markov chain implementation. The GPNP model is described in detail in Materials and Methods. In Figure 2b we plot the GPNP model histogram to indicate how well their model agrees with the data collected from the proliferating epithelia of quite distinct organisms. The most significant apparent weakness of the GPNP model is the prediction of an absence of 4-sided cells, whereas from observation about 3%−4% of cells are found to be 4-sided.

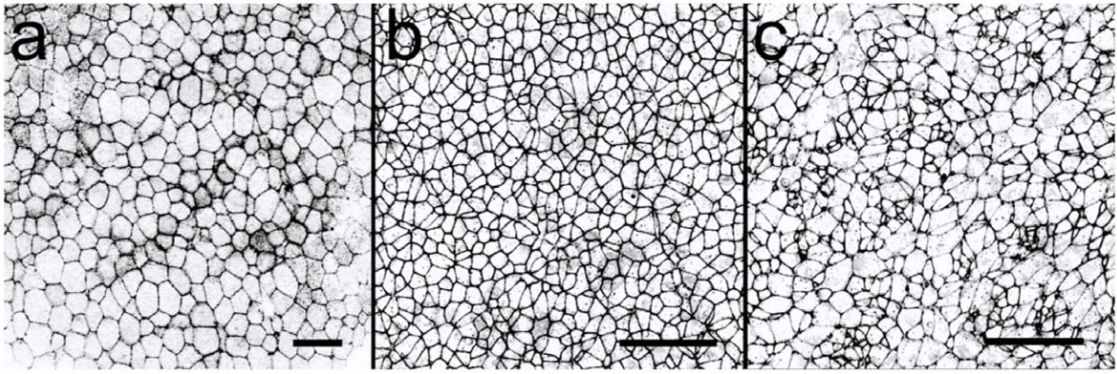

**Figure 1. High magnification images of small areas of the chick embryo.** A: stage EG XII, prior to streak formation; B: stage HH2, lateral to the streak; C: stage HH2, within the streak region. The embryo in image A is stained with rhodamine phalloidin to visualize the F actin cortex, while the embryos shown in B and C are stained with an antibody against the apically localized tight junction marker ZO-1. The areas shown are 375 × 375 $\mu m$ in A, and 187.5 × 187.5 $\mu m$ in B and C. The scale bars represent 50 $\mu m$ in all three panels.
doi:10.1371/journal.pone.0018081.g001

The primary assumptions of the GPNP model are 1) the complete absence of three-sided cells, 2) that the spindle axis defining the orientation of the division is chosen randomly for each cell, 3) that cells divide synchronously in discrete generations, and 4) that the spatial correlations between the sidedness of cells can be neglected.

Within the context of a Markov chain model we pursued two improvements by reexamining assumptions 1 and 2. Assumption 1 has been previously discussed in the supplementary information of GPNP [2], thus we keep our discussion brief. Details can be found in Materials and Methods. We reformulated the Markov chain model, allowing 3-sided cells to exist transiently, which in turn will allow a non-zero steady-state population of 4-sided cells. The model predicts about 8% of 4-sided cells, over-estimating this population by a factor of two. This error has a knock-on effect of distorting the rest of the histogram and negatively affecting the agreement with experimental data (Figure 3a).

Turning to assumption 2, the implementation of random division by GPNP is not consistent with a strictly random choice of the division axis, but, rather, imposes on the cell a particular mechanism for random axis determination. Implementing a strictly random choice of division axis, which assumes no particular cell mechanism for division, leads to a significant relative distortion of the histogram bins for 5- and 6-sided cells, again distorting the previously very good agreement with experimental data (Figure 3a). Combining the two changes to the model, i.e. allowing transient 3-sided cells and using statistically unbiased weights for the division axis only leads to worse agreement still (Figure 3a).

**Markov process models.** Given the poor results obtained with the Markov chain model, we attempt to incorporate more plausible biology, but still within the context of a non-spatial model; namely, we reexamine assumption 3. Cells in proliferating epithelia do not divide in synchronized generations, but rather, at a given time, a few widely dispersed cells will be undergoing division. We therefore attempt to bring the model closer to the observed biology by formulating cell division as an asynchronous process, using the formulation of stochastic Markov processes, in which time is now a continuous, rather than a discrete, quantity (see Materials and Methods for details). In this model, each cell has a small constant probability per unit time to undergo cell division. Thus in any given small window of time, only a small fraction of the cells in the system will be undergoing division, in accordance with experimental observation. Apart from the asynchronous division, the only other difference of implementation to the Markov chain model is the way in which new sides are distributed to neighbors of a dividing cell. In the model of GPNP, a mean field assumption is utilized (by necessity, given the non-spatial nature of the model), and each cell is given one additional side per generation, arising from the fact that on average each cell receives a side from a dividing mother cell per generation. In the Markov process model, which is also a non-spatial model, each time a mother cell divides, we choose two cells completely at random and provide each of them with an additional side. On average each cell in a starting population of $N$ cells will receive an additional side after $N$ cell divisions. An interesting by-product of using the Markov process is that even if 3-sided cells are strictly forbidden, the model will generate a non-zero fraction of 4-sided cells in the steady-state. (The technical reason for this is described in Materials and Methods.)

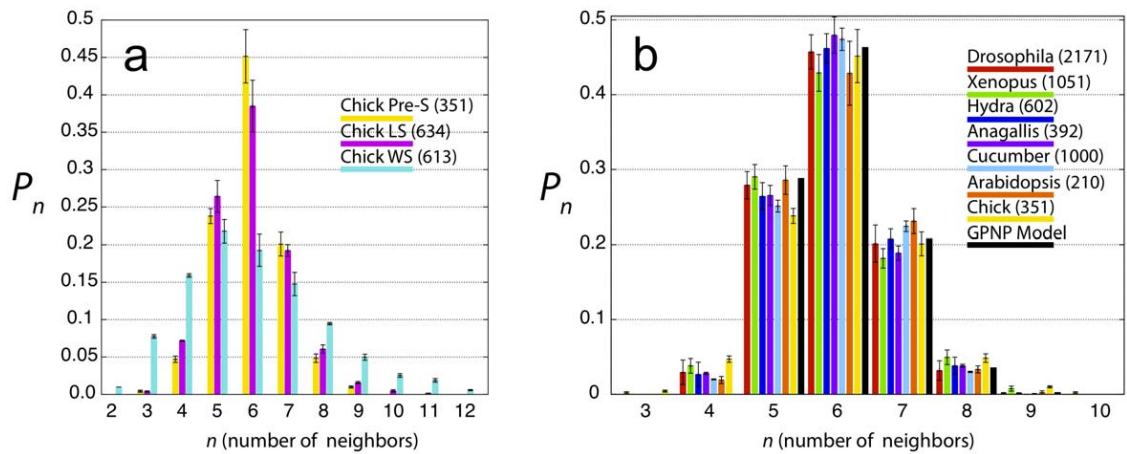

**Figure 2. Experimental CNN histograms.** A) Histograms of cell neighbor numbers for the three regions of the chick embryo shown in Figure 1; B) Comparison of the pre-streak chick histogram (yellow) with histograms reported in the literature: *Drosophila* (red), *Xenopus* (green), and *Hydra* (blue) [2]; *Anagallis* (purple) and cucumber (cyan) [7]; and *Arabidopsis* (orange) (reanalyzed using image from [6]). Also shown is the histogram of the GPNP model (black). Numbers in parentheses indicate number of cells counted in each tissue.
doi:10.1371/journal.pone.0018081.g002

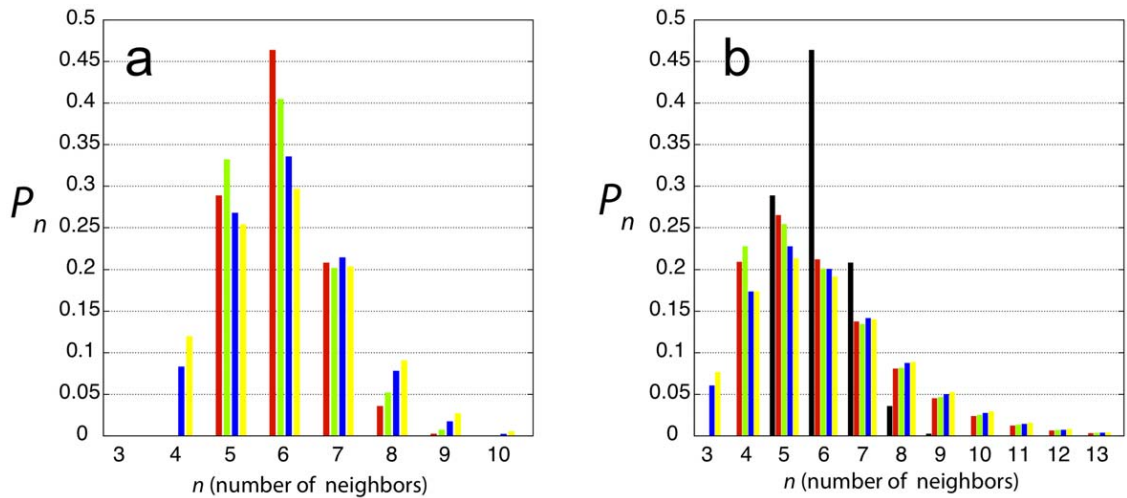

**Figure 3. Theoretical CNN histograms.** A) the histograms for the original GPNP Markov chain (MC) model (red), the MC model with adjusted statistical weights to represent a strictly random orientation of the division axis (green), the MC model allowing transient 3-sided cells thus generating a non-zero population of 4-sided cells (blue), and the MC model allowing transient 3-sided cells, *and* adjusting the statistical weights (yellow); B) the histograms for the original GPNP Markov chain model (black), and the Markov process model with four permutations of statistical weights and allowing transient 3-sided cells: GPNP weights and no 3-sided cells (red), adjusted weights and no 3-sided cells (green), GPNP weights allowing transient 3-sided cells (blue), and adjusted weights allowing transient 3-sided cells (yellow).
doi:10.1371/journal.pone.0018081.g003

The Markov process model is implemented using the Gillespie algorithm [17] and only a few seconds of CPU time are required to generate statistically precise populations of different CNN in the steady-state. As can be seen from Figure 3b, the histograms from the Markov process model are grossly distorted and bear little relation to the universal histogram. Disallowing three-sided cells, and using the statistical weights of GPNP, we find that 5-sided cells are the most common, and that cells with large numbers of sides have non-negligible statistical weight. Permuting whether or not 3-sided cells are allowed, and using the two different statistical weights has little impact on the histograms, all of which have a broad distribution for larger values of the number of neighbors, and only minor differences for smaller values.

We conclude from these results that attempts to improve the non-spatial model by adding biological realism are futile, and that the excellent agreement of the GPNP model with experimental data appears to be serendipitous. The weakness of all the models considered so far in this paper is assumption 4, namely that one can ignore the spatial correlations between cells of different sidedness. This assumption is implemented as a mean-field approximation when distributing new sides to neighbors of a dividing mother cell. In abandoning this assumption it is necessary to formulate an explicitly spatial model of the system.

**Spatially explicit simulations.** Simulating the two-dimensional polygonal projection of cells in a growing epithelial sheet is a special case of the non-trivial problem of accounting for, within computer simulations, the irregular shapes that cells can assume. Methods that have recently become available are the three-dimensional Delaunay triangulation method of Meyer-Hermann and co-workers [18], the two-dimensional vertex model [5], a dynamical variation of the vertex model [6,13], and the three-dimensional Subcellular Element Model (ScEM) [11,19]. The ScEM, in which cells are represented by coupled spatial clusters of subcellular elements, has been shown to reproduce, semi-quantitatively, the biomechanical response of cells to static and dynamical stress [20]. We report here results from our implementation of the ScEM, in a two-dimensional projection, to generate epithelial sheets through repeated cell growth and cell division. Details of the ScEM and its implementation can be found in Materials and Methods.

CNN histograms generated from our ScEM simulations show significantly better agreement with experimental data than the non-spatial models discussed above (Figure 4). Naturally, the ScEM contains several parameters, describing the mechanical properties of cells, and their dynamics. Parameters governing biomechanics (elasticity and damping of cells) can be calibrated within biologically plausible bounds using results from cell rheology experiments [20]. The histograms generated by the ScEM are relatively insensitive to most parameters. Sensitivity is found with respect to the rate of cell growth (and hence cell division). This is controlled by a parameter which determines the ease with which new subcellular elements are introduced into each cell thereby increasing its size. On varying this parameter within stability limits, we find a relatively narrow variation of histograms, as shown in Figure 4. Within this range, the ScEM produces a CNN histogram which is within the error bars of the experimental data for each bin.

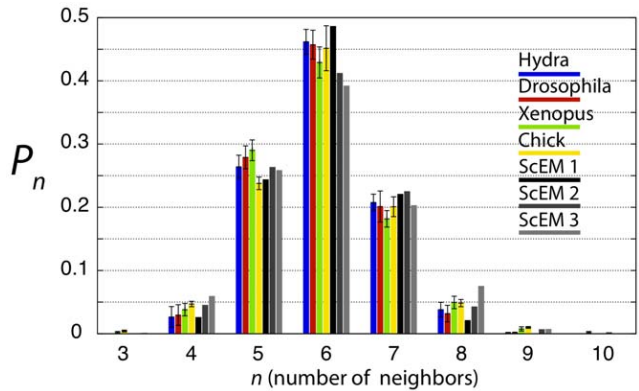

**Figure 4. Computational CNN histograms.** Histograms measured from spatial simulations of epithelial cell proliferation using the Subcellular Element Model, with three different rates of cell growth, compared with the animal subset of experimentally measured histograms described in Figure 2B.
doi:10.1371/journal.pone.0018081.g004

### In a class of its own: the within-streak histogram

The histograms for proliferating epithelia in *Hydra*, *Xenopus*, *Drosophila*, and chick (Pre-S and, to a lesser degree, LS) are remarkably similar, and appear to be examples of a universal histogram. In all these cases, the tissue dynamics is coherent, composed of local cell proliferation, and in the case of chick LS, coordinated cell migration [21]. The measured histogram for chick WS is clearly distinct, as is the tissue dynamics. In the streak region cells are ingressing to the layer below and also undergoing EMT. The histogram for WS reveals a much broader spread of cell neighbor numbers, with some cells having as many as 12 or 13 neighbors. One also sees cells with as few as 2 neighbors, i.e. cells with two convex sides.

We have argued at length above that a spatial model is required to describe the universal CNN histogram of coherent proliferating epithelia. The case of chick WS, though, is significantly different. With cells unilaterally ingressing, one might argue that such events break up spatial correlations between nearby cells. If this is the case, it would prove interesting to compare the chick WS histogram with that generated by a non-spatial model. The non-spatial model with the most plausible biological assumptions is the Markov process model described above, which accounts for the asynchronous nature of cell divisions, and which allows transient three-sided cells, and unbiased cell division statistics. On plotting the CNN histogram for this model against the chick WS histogram one finds almost perfect agreement (Figure 5). Note, there are no adjustable parameters in the Markov process model to fine tune a "goodness of fit". Given that 2-sided cells are observed (albeit in tiny numbers), we have recalculated the Markov process model allowing 2-sided cells. This results in very minor changes, and the resulting histogram provides an equally good comparison to the data.

### Discussion

To briefly summarize our results, we have analyzed histograms of cell neighbor numbers (CNN) for three different spatio-temporal regions in the early chick embryo: epiblast prior to streak formation (Pre-S), epiblast lateral to the streak (LS), and within the streak region itself (WS). Histograms of Pre-S and, to a lesser degree, LS agree well with the universal histogram measured by GPNP. The histogram for WS is very much broader. In trying to model the universal histogram we have revisited the non-spatial Markov chain model devised by GPNP. In critically analyzing their model by incorporating more plausible biology (allowing transient three-sided cells, re-examining the weights used to describe a random axis of division, and introducing a Markov process model for asynchronous division), we have consistently found significantly worse agreement with the "universal histogram". We conclude that the experiments cannot be well described by a non-spatial model, and that the agreement of the GPNP Markov chain model with experimental data appears to be serendipitous. We have turned to the Subcellular Element Model (ScEM) and produced computer-generated sheets of proliferating epithelia. Histograms measured from these sheets are in good agreement with the experimental data of GPNP, indicating that spatial correlations are a crucial component of CNN statistics for coherent proliferating embryonic epithelia. The chick WS histogram corresponds to incoherent tissue dynamics, including random ingression events, which presumably lead to a loss of local spatial correlation in the tissue. Following this intuition, we find that the chick WS histogram can be very well-described by the non-spatial Markov process model, with or without the allowance of 2-sided cells.

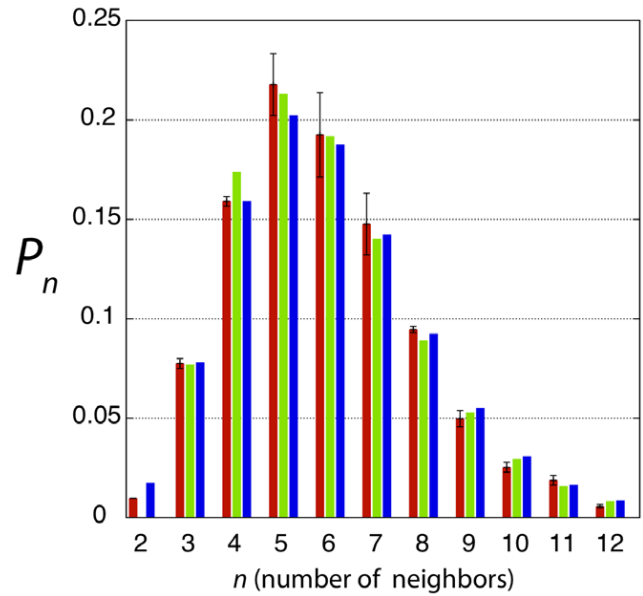

**Figure 5. CNN histograms for the primitive streak region.** Comparison of the histogram measured for the "within-streak" region of the chick embryo (red), and the histogram generated by the non-spatial Markov process model, allowing transient 3-sided cells and using the statistical weights corresponding to unbiased axis of cell division (green). Also shown is the histogram generated by the Markov process model allowing transient 2-sided cells (blue).
doi:10.1371/journal.pone.0018081.g005

A paper by Farhadifar et al. [5] also studied the CNN histogram for the *Drosophila* imaginal wing disk. Their measurements (which were made by an automated algorithm) were significantly different to those of GPNP (which were obtained "by hand"). We investigated whether this difference could be explained by the counting algorithm, rather than being due to some more subtle difference between experimental protocols or biological condi-tions. We used an automated algorithm to create CNN histograms from our ScEM data, and varied a cut-off parameter which determined whether a nearby cell was or was not in direct contact with the cell in question. With the ScEM data, just as with a pixilated micrograph, there is uncertainty involved in determining whether two given cells are neighbors. For this reason we measured a spectrum of CNN histograms for a range of interaction cutoffs $1 \leq r_c/r_0 \leq 2$, where $r_0$ is the diameter of a subcellular element and $r_c$ is the maximum distance for which neighboring elements from different cells are considered to define a cell-cell contact. This issue is closely related to the problem of resolving apparent four-way junctions (see Materials and Methods for more information). A similar spectrum of CNN histograms was measured in the *Drosophila* wing disc by Farhadifar et al. [5] (supplementary material). These authors used a different parameter: cell-cell boundary length. Any two proximate cells were considered neighbors if their boundary length was greater than a percentage of the average cell-cell boundary length. If we consider just the *Drosophila* data by GPNP in Figure 6, we see that this histogram is more sharply peaked than that of Farhadifar et al. [5]. In comparison the Farhadifar et al. [5] histogram is biased more towards lower cell neighbors and the peak has ∼6% less 6-neighbored cells than the GPNP data. We found that on varying the element-element interaction cut-off parameter $r_c$, our measured histograms interpolated smoothly between those obtained by Farhadifar et al. and GPNP, thereby providing a simple possible explanation for the observed differences (Figure 6).

As mentioned in the Introduction, there have been several recent papers reporting spatial models of epithelial topology. Patel et al [7] extended the Markov chain model by focusing on tissue topology, while neglecting tissue mechanics. They studied different algorithms for choosing the cell division axis. They found two choices which provide a reasonable match to the universal histogram, and provided arguments to connect these choices to different cell division mechanisms in plants (*Anagallis* and *Cucumis*) and animals (*Drosophila*). However, the morphology of the tissues created using their model are highly "splintered", bearing no resemblance to actual epithelia. An alternative modeling approach, based on a vertex model, was pursued by three different groups [5,12,14]. In this approach cells are assumed to be precisely polygonal, so that they are completely defined by their vertices. Images of the chick epiblast (Figure 1) indicate that cell boundaries often have significant deviations from a straight line, but a vertex model is a reasonable zeroth-order approximation of cell shapes. A global energy function is written down for the vertices, accounting for bulk cell compression, line tension, and contractility. Various algorithms for cell growth and division are then implemented, and after each cell event, the global energy is minimized. One weakness of this model is that global energy minimization can lead to the collapse of vertices to a point, which in effect eliminates a cell. This process is referred to as "apoptosis" [5]. This defect can be ameliorated considerably by allowing active cell rearrangements to occur when cell shrinkage threatens to eliminate a cell [12]. The original implementation of this model by Farhadifar et al [5] provided a useful phase diagram relating different tissue topology phenotypes with variations in the cell-scale elastic parameters. However, histograms arising from the model differed markedly from the universal histogram of GPNP (generally the peak of the published distributions occurred for 5-sided cells). Significant improvement was achieved by Aegerter-Wilmsen et al [12] by implementing a cell growth rate that increases with apical cell surface area, and by allowing for small cell growth increments in each dynamical cycle, rather than allowing each cell to successively double in size and then divide while holding the rest of the tissue fixed, as in the previous implementation [5]. They also measured CNN histograms for the subset of mitotic cells, and found good agreement with experimental data, providing additional credibility to their model, and a strong case that mechanical regulation of growth rate is important. The ScEM implementation presented in the current paper does not have a cell-size dependence on growth rate. As described in Materials and Methods, cell growth (i.e. adding a new element to the core of a cell) occurs adiabatically relative to time-scales of element-element equilibration, and thus cell densities are uniform. Naveed et al [14] also used the vertex model, and studied two different choices for determining the cell division axis. They found that selecting the division axis to originate from the longest side of the cell gave good agreement with the universal histogram. Sahlin et al [6,13] introduced a different algorithm based on a vertex model, in which vertices were coupled by overdamped springs, and had uniform growth. Their primary interest was in CNN histograms for plant tissues. They studied various rules for division, and were able to obtain reasonably good agreement with the CNN histogram for *Arabidopsis* by using rules which tended to produce isotropic and equally-sized daughter cells. Clearly, more work will be required on spatial modeling to identify common ground between these various spatial models and thus determine the key cell biological variables which underlie the universal CNN histogram. It may well be that purely two-dimensional models do not contain enough biological detail to provide a compelling resolution to this question. Cell division in epithelial sheets is a truly three dimensional event, as described below.

A recent paper [22] explained the universal histogram in terms of an energy minimization argument, reminiscent of work on soap films, and other non-biological cellular structures. Key to this argument is the existence of an order parameter, the reduced area $a$ (defined as $4\pi$ multiplied by the ratio of cross-sectional area to the square of the perimeter). Different tissues are presupposed to be each characterized with a particular value of $a$, and therefore have different morphologies (and hence different CNN histograms). We measured the reduced areas for cells in the chick Pre-S dataset and found that within this one system, there was a very significant variation in the reduced area for cells (see File S1 for details). This contradicts the assumption of using the reduced area as a robust order parameter, at least for the chick epiblast system.

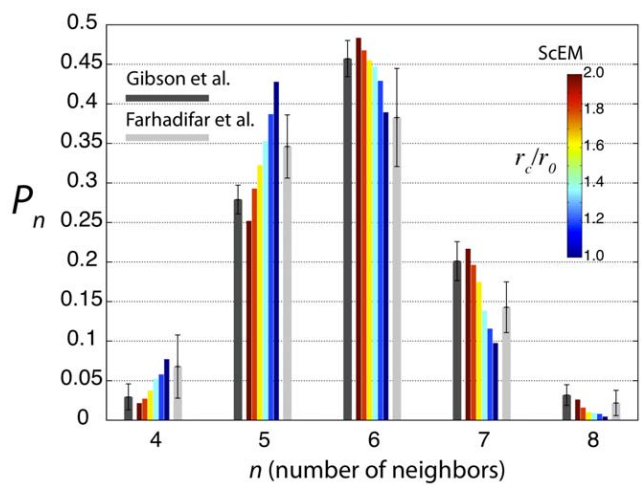

**Figure 6. Dependence of CNN histograms on neighbor criterion.** Histograms for the *Drosophila* imaginal wing disk from Farhadifar et al. [5] and GPNP. Also shown are five histograms from ScEM simulations, in which the cut-off criterion $r_c/r_0$ for what constitutes a neighboring cell is smoothly varied.
doi:10.1371/journal.pone.0018081.g006

There are several finer points which bear discussion. First among these is the ScEM implementation of cell division. We have used an algorithm in which cells choose their division axis perpendicular to the long axis of the cell, not unlike the choice favored in the recent paper by Naveed et al [14]. This rule is also similar to Errera's rule [23] from botany, namely that plant cells will tend to divide such that the shortest line is used in the plane of the cell [6]. Within the confines of a purely two-dimensional simulation, this choice is favorable as it allows cells to retain a reasonable degree of isotropy. Cell divisions chosen purely at random tend to lead to highly anisotropic cells, which, despite their propensity to reduce surface area (or rather, peripheral length in two dimensions), are unable to round up over the time scales of proliferation. As already mentioned, an extreme example of this type of division can be seen in a recent spatial model in which cell mechanics is neglected [7]. It is important to emphasize that our chosen algorithm still ensures that the cell division axes sampled over the entire cell population are uniformly distributed (i.e. isotropic).

It has also become clear during our investigations that it is not unambiguous how to assign the number of neighbors to cells. Epithelial cells are generally columnar in shape and the number of neighboring cells contacting a given cell at the apical side at the level of the adherens junctions may not be exactly the same as the number of cells contacting this cell at the basal side. From our experience this is the case in the chick embryo and is likely to hold true for other organisms such as *Drosophila* as well. In the case of the chick embryo epiblast, cells in M phase contract in the apical direction and round up, a process which is coupled to movement of the nuclei in the apical direction. It is to be expected that the cells keep some contact with the basal lamina and that this informs the mitotic cell of its position in space, and allows the cell to orient its spindle and contractile ring such that both daughter cells remain in the epiblast. This pattern of division suggests that the plane and position of cell division is controlled primarily by mechanical constraints. It is not known how this is achieved but it makes it likely that more realistic models that try to predict the number of neighbors must take these mechanical considerations into account. Spatial models which can accommodate cell mechanics in 3D, such as the ScEM, will be of significant value in this exploration, especially when coupled with 3D live imaging of cells in dividing epithelia. Extensions to the ScEM, necessary to describe the active processes of cell rounding and division, are non-trivial and beyond the scope of this paper.

The fact that CNN histograms for coherent and incoherent cell dynamics in the chick embryo are significantly different indicates that an analysis of neighbor numbers on a single image may provide insight into the underlying dynamics of the system. This may have potential value for histological examination of tissue biopsies, for example identifying from a fixed sample whether or not a process such as metastasis is occurring. However, a recent quantitative analysis relating cell behaviors and tissue dynamics revealed that embryonic tissues undergo numerous dynamical tissue phenotypes simultaneously [24]. This indicates that the relation between coherent and incoherent tissue dynamics and tissue topology may be more complicated than would appear from the analysis presented here.

## Materials and Methods

### Experiments

Fertile eggs (High Sex X Rhode Island Red) were obtained from Winter Farm, Thirplow, Herts, UK. The embryos were cultured in EC culture [25] and incubated for 1–12 hours at 38°C in a humid incubator. The embryos were fixed in 4% PFA in PBS pH 7.4 for 2 hours on ice, followed by washing 3 times for 30 min with PBST (PBS containing 0.1% Tween20). F-actin staining was performed by incubating the embryos in PBS containing 0.02 ugr/ml TRITC conjugated phalloidin (Sigma, P1951). For antibody staining the embryos were pre-incubated with 0.3% H2O2 in PBS for 1 hour, washed once by PBST followed by immersion overnight at 4°C in a blocking solution (PBST, 2% BSA, 10% normal goat serum) containing a 1:100 dilution of an antibody against ZO1(Invitrogen Cat No: 40-2200). After washing three times in PBST the embryos were incubated with HRP conjugated anti-rabbit antibody (Promega, W401B) 1:1000 dilution in blocking solution overnight at 4°C. This was followed by washing twice with PBST and development with Alexa-Fluor 488 Tyramide488 Signal Amplification Kit (Molecular Probes, Inc) for 30 minutes at room temperature.

### Markov chain

The assumption that cells divide synchronously in discrete generations allows the system to be cast, quite elegantly, as a Markov chain [2]. The fraction of cells with various sidedness at the next generation can be expressed in terms of the fractions at the current generation using a matrix of transition probabilities which account for how random divisions connect mother cells of sidedness $n$ to daughter cells of sidedness $l$ and $m$. Note, simple geometry dictates that $n = l + m - 4$. Given assumption 1, i.e. that $n \geq 4$, we have $4 \leq l \leq n$ and similarly for $m$.

Following GPNP we decompose the transition matrix into two successive matrices, the first accounting for the sidedness of daughter cells created by a given mother cell dividing, the second accounting for the extra sides picked up by cells neighboring a mother cell when it divides. We define by the column vector $p = (p_4, p_5, \cdots, p_n, \cdots)^T$ the fractions of cells with different sidedness at the current generation. For the first part of the transition matrix we write the intermediate state as $p_m^* = T_{m,n}^{(0)} p_n$. Defining the combinatorial symbol in the usual way, i.e. $C_j^i = i!/j!(i-j)!$, GPNP write

$$T_{m,n}^{(0)} = C_{m-4}^{n-4}/2^{n-4}. \quad (1)$$

This form arises from the following argument. Assuming that a mother cell of $n$ sides (and hence $n$ vertices) divides along a random orientation, one must compute the probability that the division axis separates $k$ vertices on one side and $n-k$ vertices on the other (which would lead to two daughter cells of sidedness $l = k+2$ and $m = n-k+2$). Since each daughter cell must have at least four sides, there must be at least two vertices on each side of the division axis. GPNP proceed to take 2 vertices from the total of $n$ and place them on one side, two more and place them on the other side, leaving $n-4$ which are to be distributed randomly to the two sides. This gives rise to the binomial form written above. The second part of the transition matrix arises as follows. Let us denote the number of mother cells at the current generation by $N$. After each cell has divided, the total number of cells has increased to $2N$. Each division will have led to two additional sides being provided to the daughter cells, *and* to two neighboring cells each being provided with one additional side. Thus $2N$ sides are created that have to be distributed to neighboring cells. Thus, *on average*, each daughter cell of the next generation picks up an additional side from being a neighbor of a dividing cell. Ignoring spatial correlations between cells, in the spirit of a mean field approximation, GPNP assume that each cell *actually* picks up an additional side. Denoting by $p'$ the state of the new generation, we have $p'_m = S_{m,n} p^*_n$ where $S_{m,n} = \delta_{m,n+1}$. Combining these two transition matrices, we have the transition matrix $U$ connecting successive generations as $p'_m = U^{(0)}_{m,n} p_n$, where

$$U^{(0)}_{m,n} = C^{n-4}_{m-5}/2^{n-4}, \qquad (2)$$

where entries in U are assumed to be zero unless $n \geq 4$ and $5 \leq m \leq n+1$. Iteration of this map leads to a steady-state for p.

**Critique: allowing transient three-sided cells.** It can be seen from the form of U that although 4-sided cells are allowed, there is no entry in the transition matrix U which can create them. Hence, the fraction of 4-sided cells decreases monotonically under iteration to a steady-state value of zero, which is not compatible with the experimental observations. This prompts one to allow three-sided daughter cells to be created transiently by the first part of the transition matrix, since they will be converted to four-sided and five-sided cells by the transition matrix. In this case, when a mother cell of sidedness $n$ divides to create two daughter cells of sidedness $l$ and $m$, we still have $n = l + m - 4$, but now $3 \leq l \leq n+1$ and similarly for $m$. Following exactly the same logic as before we have $p^*_m = T^{(1)}_{m,n} p_n$ and $p'_m = U^{(1)}_{m,n} p_n$, where

$$T^{(1)}_{m,n} = C^{n-2}_{m-3}/2^{n-2}, \qquad (3)$$

and

$$U^{(1)}_{m,n} = C^{n-2}_{m-4}/2^{n-2}. \qquad (4)$$

The entries for U are assumed to be zero unless $n \geq 3$ and $4 \leq m \leq n+2$. One sees from this map that although 3-sided cells are allowed in principle, the fraction decreases monotonically to zero at the steady-state, in accordance with the negligible numbers of 3-sided cells observed in experiments, but that there will remain a non-zero fraction of 4-sided cells, also in accordance with observation.

**Critique: conditional probabilities and cell division.** The argument leading to the form for $T^{(0)}_{m,n}$ given in Eq.(1) contains a subtle bias concerning conditional probabilities, and is not compatible with assumption 2, namely that the division axis is chosen *completely at random*. The algorithm of GPNP is to take 2 vertices and place them on one side of the axis, 2 more vertices and place them on the other side of the axis, and then to randomly distribute the remaining $n-4$ vertices. This algorithm is not unique. To illustrate this, consider an alternative, and admittedly awkward, algorithm. First randomly distribute $n-2$ vertices. Then, with 2 vertices remaining there are three possibilities: i) if one side has no vertices, provide that side with the two vertices, ii) if one side has only one vertex, randomly distribute one vertex, and if that side still only has one vertex, then provide it with the final vertex, otherwise randomly distribute the final vertex, and iii) if each side already has at least two vertices, randomly distribute the remaining two vertices. This algorithm for distributing vertices will lead to a significantly different distribution of daughter cells than the one chosen by GPNP. There are many other algorithms one can concoct that all have different final distributions of daughter cells. One thus sees that the algorithm chosen by GPNP is ad hoc, unless one believes that the cell, in dividing, causally ensures that two vertices are on the left, two are on the right, and then makes a random choice of orientation among the remaining vertices. To truly capture the assumption of random division, one must assume that the cell chooses a division axis purely at random, and then discards outcomes that are not consistent with the constraint of at least two sides per daughter cell. In this case, we have simply the binomial distribution, with a corrected normalization accounting for the fact that $2 + 2n$ outcomes are discounted:

$$T^{(2)}_{m,n} = C^n_{m-2}/(2^n - 2 - 2n). \qquad (5)$$

This leads to the final transition matrix

$$U^{(2)}_{m,n} = C^n_{m-3}/(2^n - 2 - 2n), \qquad (6)$$

where entries for U are assumed to be zero unless $n \geq 4$ and $5 \leq m \leq n+1$.

In the case where transient three sided cells are allowed, we again allow purely random division and discount the 2 outcomes per division in which one side does not have at least one vertex, giving

$$T^{(3)}_{m,n} = C^n_{m-2}/(2^n - 2), \qquad (7)$$

and

$$U^{(3)}_{m,n} = C^n_{m-3}/(2^n - 2), \qquad (8)$$

where entries for U are assumed to be zero unless $n \geq 3$ and $4 \leq m \leq n+2$.

### Stochastic process description

We describe the state of the epithelium at some arbitrary time $t$ by the probability distribution $P(N_3, N_4, \ldots, t)$, where $N_m$ is the number of cells with $m$ sides. We assume that in the time interval $(t, t + \delta t)$, each cell has a fixed probability $\kappa \delta t$ to divide. Given we are only keeping track of the numbers of cells of different sidedness, the fundamental transition rate in this stochastic process describes the transition of a cell with $n$ sides dividing into two daughter cells with $l$ and $m$ sides respectively. Denoting this transition rate by $W^{(q)}_{m,n}(N_n)$ we have

$$W^{(q)}_{m,n}(N_n) = \kappa N_n T^{(q)}_{m,n}, \qquad (9)$$

where the matrices $\mathbf{T}^{(q)}$ are identical to those derived above for the Markov chain model, and the index $q = 0, 1, 2, 3$ indicates whether or not 3-sided cells are allowed, and what type of binomial weights are being used to decide the probability of a particular division axis.

Two of the neighbors of the mother cell will obtain an additional side from the mother cell's division. In common with GPNP, we ignore spatial correlations and do not explicitly keep track of the sidedness of cells neighboring a given cell. In the spirit of a mean field approximation, after a given cell division process, we randomly select two cells and give each an additional side. This is equivalent, on average, to the mean field approximation of GPNP in which each cell in the population is given an additional side after one complete round of synchronous cell division. Thus, a single cell division, of a mother with $n$ sides yielding daughter cells with $l$ sides and $m$ sides, can be described by the following set of transitions:

$$N_n \to N_n - 1,$$

$$N_m \rightarrow N_m + 1,$$

$$N_l \rightarrow N_l + 1,$$

$$N_i \rightarrow N_i - 1,$$

$$N_{i+1} \rightarrow N_{i+1} + 1,$$

$$N_j \rightarrow N_j - 1,$$

$$N_{j+1} \rightarrow N_{j+1} + 1.$$

This set of transitions occurs as a block with a rate per unit time of $W_{m,n}^{(q)}$. The sidedness indices $i$ and $j$ are chosen randomly, weighted appropriately such that any cell in the sheet has an equal probability of gaining an additional side through being a neighbor of the currently dividing cell. Note, a positive advantage of the Markov process model is that 4-sided cells will have a non-zero population in the steady-state even if 3-sided cells are strictly forbidden. The reason is as follows. In the Markov chain model, the second part of the transition matrix raises the sidedness of all cells by one side. Thus there is no way to generate new four-sided cells. In the Markov process model, because each event involves a single mother cell, and new sides are distributed completely at random, a non-zero fraction of 4-sided daughter cells that are generated by cell division will survive in the steady-state population.

We have implemented this stochastic process using the Gillespie algorithm [17], which efficiently generates statistically exact realizations. In a few seconds on a single processor the algorithm can generate a single realization comprising a population with millions of cells. (Note: the algorithm only keeps track of the number of cells within each sidedness class, which are the stochastic variables in the non-spatial stochastic process defined above.) Such large populations become rapidly self-averaging, and it is straightforward to read off the cell-sidedness histogram by following the relative fractions of cells in the different sidedness classes within one realization. These relative fractions rapidly converge to the quasi-steady-state values.

## Spatially explicit simulation

The Subcellular Element Model (ScEM) was used to grow a continuous sheet of cells in two dimensions. The ScEM allows the simulation of large cell aggregates in a grid-free environment [11]. Each cell is modeled as a cluster of visco-elastically coupled elements, thereby allowing emergent cell shape dynamics. Cell-level mechanics predicted by the ScEM is in good agreement with experiments on cell rheology [20].

Here we describe implementing the ScEM in two dimensions in order to grow an epithelial-like sheet. For computational efficiency, we seed each simulation with an array of 37 cells, each composed of 128 subcellular elements. A given cell grows through a process of the random addition of elements to the cell core. As a process of regulating growth, the algorithm is as follows [19]. At each time step we allow a subset of elements (i.e. those in the cell core) to attempt a replication process with a small probability. For a given element $\alpha$ at a position $\mathbf{r}_\alpha$ we randomly select a point $\mathbf{r}'_a$ a distance $r_0$ from element $\alpha$, where $r_0$ is the diameter of an element. If this point is sufficiently far from neighboring elements $\beta$ (meaning that $|\mathbf{r}_{\alpha'} - \mathbf{r}_\beta| > d_{min}$) then a new element $\alpha'$ is placed at that point. Once the cell doubles in size (meaning that the number of elements doubles through replication), the cell splits evenly into two daughter cells of approximately 128 elements each. Note, element replication occurs with a small probability to ensure that local element-element mechanical equilibrium is not strongly perturbed. New elements are introduced adiabatically, such that cell densities are uniform throughout the tissue. In the real embryonic epithelium, cell division proceeds through a complex sequence of columnar to spherical to columnar morphological transitions. We do not attempt to model this process in the current work. We use instead a simple algorithm to determine the axis of cell division; namely we determine the geometric long axis of the cell, and divide perpendicular to this. This maintains an epithelial sheet with roughly isotropic polygonal cells. Choosing a random axis of division (random both in absolute space and relative to the long axis of the cell) yields cell morphologies which are increasingly polarized ("splintered") as proliferation continues. Similar computer-generated morphologies have been reported recently [6,7].

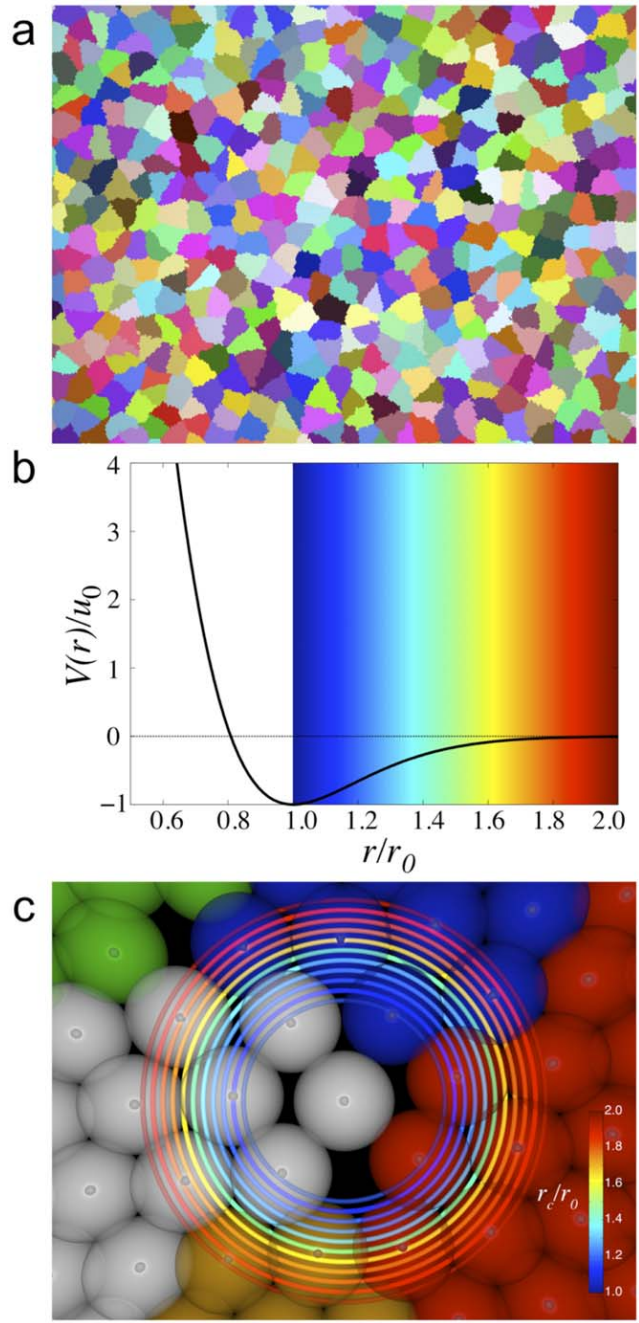

Figure 7. ScEM simulations. (a) An example of cell morphology of a cell sheet grown in two dimensions using the ScEM with 128 to 256 elements per cell. Cells are colored randomly in order to distinguish cells. (b) The scaled intercellular interaction potential between elements. Beyond the equilibrium distance $r_0$ there are decaying interactions between elements. This region is outlined with a color gradient. (c) High magnification of elements of several neighboring cells. Cells are distinguished by different colors. Demonstrated here are the different cut-off ranges for which two elements from different cells are considered to be interacting strongly enough that their associated cells are neighbors. The cut-off range is $r_c$, and $r_0$ is the diameter of an element and/or the equilibrium distance of the interaction potential shown in (b).
doi:10.1371/journal.pone.0018081.g007

Cell proliferation is allowed to continue until the system size is large enough to obtain good statistics for cell neighbor counting: about 1000 to 1500 cells (Figure 7a). The viscoelastic properties of cells were chosen so that the bulk elastic modulus of a single cell was of order 1000 Pa. Viscosity was computed so that the relaxation dynamics of the cell in response to a small perturbation was of order 1 second [26]. Methods for calibrating these values are discussed in our previous work [20]. Cell-cell adhesion, as measured by the force per unit area to dissociate two cells, was set to be approximately 250 Pa.

For a given cell, in order to determine whether a proximate cell is indeed a neighbor, we consider cell-cell interactions at the level of the subcellular elements; in particular, by considering the short-ranged element-element interaction potentials. Subcellular elements have a linear dimension of typically about 1 micron (if 128 elements are used to represent the cross-section of a cell), and so are significantly more coarse-grained than protein complexes responsible for binding cells together in an epithelium, the existence of which unambiguously defines the cells in question as neighbors. With this coarse-graining, there is uncertainty involved in determining whether two nearby cells are indeed neighbors. This same issue arises in experimental determination of neighbors, obtained by analyzing pixelated micrographs of cells. Because of these uncertainties, we extensively analyzed the sensitivity of CNN histograms as a result of varying our criteria for which two cells are defined as neighbors. The criteria we used were *proximity* and *number of subcellular element interactions*. Proximity is defined in terms of a cut-off distance $r_c$, which is the maximum distance for which neighboring elements from different cells are considered to define a cell-cell contact. As shown in Figure 7b, beyond the equilibrium distance of the potential well ($r_c \geq r_0$), the colored gradient outlines the range for which element-element interactions are definite, but cell-cell neighboring relationships could be considered uncertain. For the second criteria, it is not clear within the community whether one node or two or more boundary cell-cell interactions constitute a cell-cell neighboring relationship [5]. For this reason, our analysis entertains both cases. Further, we measure a spectrum of CNN histograms for a range of interaction cut-offs $1.0 \leq r_c/r_0 \leq 2.0$.

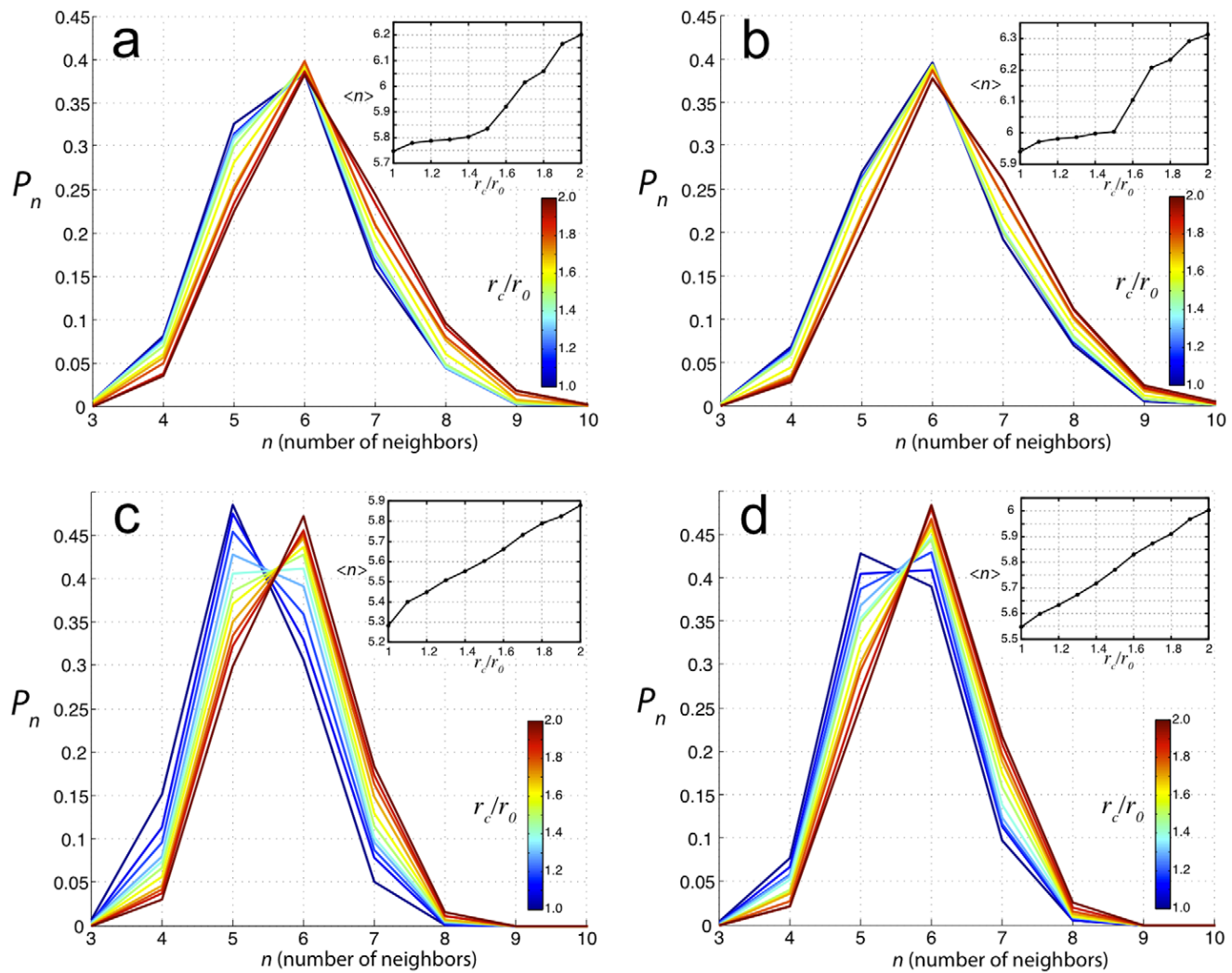

**Figure 8. Neighbor criterion investigated using the ScEM.** (a) CNN histograms of ScEM simulations in which any two cells are considered neighbors if they have at least 2 elements in contact within a range cut-off of $r_c$; $r_0$ is the constant diameter of an element. (b) CNN histograms of the same ScEM simulation as in (a) except any two cells are considered neighbors if they have at least 1 element in contact within a range cut-off of $r_c$; $r_0$ is the constant diameter of an element. Inserts shows $\langle n \rangle$ for a corresponding cut-off. (c) and (d) are the same analysis but for different simulation parameters which vary the rate of proliferation.
doi:10.1371/journal.pone.0018081.g008

In Figure 8 results from two different simulations are shown in panels (a,b) and (c,d) respectively, which correspond to two different values of the growth parameter $d_{min}$: $0.54 r_0$ for upper panels and $0.56 r_0$ for lower panels. The parameter $d_{min}$ is constrained to this range of values, due to matching biologically plausible rates of proliferation to the intrinsic time of mechanical equilibration. Increasing $d_{min}$ effectively decreases the rate of proliferation by increasing the spatial sensitivity of placing new elements. Note that a higher value of $d_{min}$, and thus a more sensitive criterion for element placement, leads to sharper histograms. The left panels (a,c) show CNN histograms using the condition that cell-cell contact is defined by just one common element-element interaction. The right panels (b,d) show CNN histograms using the condition that cell-cell contact is defined by at least two common element-element interactions. For all panels, we can see that changing $r_c/r_0$ from 1.0 to 2.0 shifts the histograms from lower to higher CNNs respectively. Inserts show $\langle n \rangle$ for a corresponding cut-off. For tightly packed cells in epithelial-like tissues, $\langle n \rangle$ is usually very close to 6. For this reason, we assume that the most accurate histogram describing our simulated tissues will be that having a value of $\langle n \rangle$ closest to 6.

## Supporting Information

**File S1** Further details on data acquisition and analysis, and a more detailed discussion of the reduced area concept. Includes three additional figures.
(PDF)

## Acknowledgments

The authors gratefully acknowledge Jayanth Banavar for helpful comments on the manuscript, and the 2007–08 program on Mathematics of Molecular and Cellular Biology, organized by the Institute for Mathematics and its Applications, University of Minnesota, where this project was initiated.

## Author Contributions

Conceived and designed the experiments: MC CJW. Performed the experiments: MC CJW. Analyzed the data: SS CJW TJN. Wrote the paper: SS CJW TJN. Theoretical and computational modeling: SS TJN.

## References

1. Gong Y, Mo C, Fraser SE (2004) Planar cell polarity signalling controls cell division orientation during zebrafish gastrulation. Nature 430: 689–693.
2. Gibson MC, Patel AB, Nagpal R, Perrimon N (2006) The emergence of geometric order in proliferating metazoan epithelia. Nature 442: 1038–1041.
3. Lewis FT (1926) The effect of cell division on the shape and size of hexagonal cells. Anatomical Records 33: 331–355.
4. Lewis FT (1928) The correlation between cell division and the shapes and sizes of prismatic cells in the epidermis of cucumis. Anatomical Records 38: 341–376.
5. Farhadifar R, Röper JC, Aigouy B, Eaton S, Jülicher F (2007) The influence of cell mechanics, cell-cell interactions, and proliferation on epithelia packing. Curr Biol 17: 2095–2104.
6. Sahlin P, Hamant O, Jönsson H in *Complex Sciences, Lecture Notes of the Institute for Computer Sciences, Social Informatics and Telecommunications Engineering*, Vol. 4 (Springer Berlin). pp 971–979.
7. Patel AB, Gibson WT, Gibson MC, Nagpal R (2009) Modeling and inferring cleavage patterns in proliferating epithelia. PLoS Comp Biol 5: e1000412.
8. Chickarmane V, Roeder AH, Tarr PT, Cunha A, Tobin C, et al. (2010) Computational morphodynamics: a modeling framework to understand plant growth. Annu Rev Plant Biol 61: 65–87.
9. Stern CD (2004) Gastrulation: from Cells to Embryos, (Cold Spring Harbor Laboratory Press, New York).
10. Kadanoff LP (2002) Statistical Physics, (World Scientific, Singapore).
11. Newman TJ (2005) Modeling multi-cellular systems using sub-cellular elements. Math Biosci Eng 2: 611–622.
12. Aegerter-Wilmsen T, Smith AC, Christen AJ, Aegerter CM, Hafen E, et al. (2010) Exploring the effects of mechanical feedback on epithelial topology. Development 137: 499–506.
13. Sahlin P, Jönsson H (2010) A modeling study on how cell division affects properties of epithelial tissues under isotropic growth. PLoS ONE 5: e11750.
14. Naveed H, Li Y, Kachalo S, Liang J (2010) Geometric order in proliferating epithelia: impact of rearrangements and cleavage plane orientation. Conf Proc IEEE Eng Med Biol Soc 2010. pp 3808–3811.
15. Eyal-Giladi H, Kochav S (1976) From cleavage to primitive streak formation: a complementary normal table and a new look at the first stages of development of the chick. I General morphology, Dev Biol 49: 321–337.
16. Hamburger V, Hamilton HL (1951) A series of normal stages in the development of the chick embryo. J Morphol 88: 49–92.
17. Gillespie DT (1976) A general method for numerically simulating the stochastic time evolution of coupled chemical reactions. J Comp Phys 22: 403–434.
18. Schaller G, Meyer-Hermann M (2004) Kinetic and dynamic Delaunay tetrahedralizations in three dimensions. Comp Phys Comm 162: 9–23.
19. Newman TJ (2007) in Single Cell Based Models in Biology and Medicine, Anderson A, Chaplain M, Rejniak K, eds. (Birkhäuser, Basal). pp 221–239.
20. Sandersius SA, Newman TJ (2008) Modeling cell rheology with the Subcellular Element Model. Phys Biol 5: 015002. 13 p.
21. Chuai M, Weijer CJ (2008) The mechanisms underlying primitive streak formation in the chick embryo. Curr Top Dev Biol 81: 135–156.
22. Hocevar A, Ziherl P (2009) Degenerate polygonal tilings in simple animal tissues. Phys Rev E 80: 011904. 7 p.
23. Errera L (1888) Uber zellformen und siefenblasen. Botanisches Centralblatt 34: 395399.
24. Blanchard GB, Kabla AJ, Schultz NL, Butler LC, Sanson B, et al. (2009) Tissue tectonics: morphogenetic strain rates, cell shape change and intercalation. Nature Methods 6: 458–464.
25. Chapman SC, Collignon J, Schoenwolf GC, Lumsden A (2001) Improved method for chick whole-embryo culture using a filter paper carrier. Dev Dyn 220: 284–289.
26. Wottawah F, Schinkinger S, Lincoln B, Ananthkrishnan R, Romeyke M, et al. (2005) Optical rheology of biological cells. Phys Rev Lett 94: 098103. 4 p.